# AI-Assisted Hyperspectral Interferometry and Single-Cell Dispersion Imaging

**KAMYAR BEHROUZI,[1,2] TANVEER AHMED SIDDIQUE,[3] MEGAN TENG,[1,2], WALID REDJEM,[3] LIWEI LIN,[1,2,*] AND BOUBACAR KANTE[3,4,*]**

[1]*Department of Mechanical Engineering, University of California Berkeley, Berkeley, CA, 94720, USA*
[2]*Berkeley Sensor and Actuator Center (BSAC), Berkeley, CA, 94720, USA*
[3]*Department of Electrical Engineering and Computer Science, University of California Berkeley, Berkeley, CA, 94720, USA*
[4]*Material Science Division, Lawrence Berkeley National Laboratory, Berkeley, CA, USA*

**lwlin@berkeley.edu*
**bkante@berkeley.edu*

**Abstract:** Interferometry techniques are essential for extracting phase information from optical systems, enabling precise measurements of dispersion and highly sensitive detection of perturbations. While phase sensing offers enhanced sensitivity compared to conventional spectroscopy methods, this sensitivity often makes systems more vulnerable to external factors such as vibrations, introducing instability and noise. In this work, we demonstrate a broadband and AI-enhanced interferometry method, denoted general polarization common-path interferometry (GPCPI), that relaxes the polarization constraints of traditional common-path interferometry. The polarization decoupling feature enables simultaneous amplitude and phase measurements supplemented with deep neural autoencoders to detect phase anomalies in the spectrum through the analysis of second order derivative mapping of the phase profile, enhancing the accuracy of broadband phase measurements. The approach enables an order of magnitude improvement in phase stability compared to state-of-the-art interferometry techniques, leading to higher accuracy in phase sensing. Plasmonic metasurface phase sensing and hyperspectral single-cell dispersion imaging demonstrate the capability and sensitivity of the method over conventional spectroscopy. Our own adopted version of deep learning model, ConvNeXt V2, enables single-shot and real-time tracking of phase variation with minimized noise. Interference fringes affected by the cell-cultured samples reveal the fingerprints of the normal (CCD-32Sk) vs cancerous (COLO-829) skin cells, enabling cell classification and disease diagnosis at single-cell level through hyperspectral dispersion imaging. The proposed interferometry technique offers a reliable, compact, and stable solution for broadband phase measurements and single-cell dispersion imaging for applications in metrology, molecular diagnostics, drug discovery, and quantum sensing.

## 1. Introduction

Measuring light-matter interaction offers extensive insights into material properties [1,2]. Measurements can be based on intensity using spectroscopy techniques [3–5] or more advanced methods incorporating polarization [6–8] and/or phase [9–11]. Although the materials spectra provides rich information about chemical bonds [12,13] or resonating modes [14], it does not capture the polarization response [15] or the dispersive properties [16–19]. Phase measurements combined with spectroscopy techniques allows for the characterization of the complex response of materials and devices [11,16,20]. Phase measurements can be directly employed to detect small perturbations in an optical system, such as thickness variations or the presence of micro/nanoparticles [21–29].

Conventional phase measurement techniques, such as Michelson interferometry (MI) [30], are prone to vibrations, thermal variations, and environmental perturbations [31]. To mitigate these effects, the common-path interferometer (CPI) was introduced, reducing the two paths of the interferometer (sensing and reference) to a single path. The CPI has mostly been based on preferential interaction of a sample with one polarization of light or applied to samples with special optical mode design [21,22]. In **Fig. 1**, we propose to combine MI and CPI, denoted general polarization CPI (GPCPI), that is simultaneously stable and independent of input polarization. GPCPI eliminates requirements on the sample's polarization properties by generating reference and sensing arms immediately before the sample using a Wollaston prism. The polarization decoupling feature of GPCPI enables the simultaneous measurement of samples transmittance and phase spectra, using either a polarization sensitive camera or multiple cameras with linear polarizers.

Besides stability and polarization independence, the spectral bandwidth of the measurements is a key factor in interferometry as it provides information about the dispersive properties of materials [16–19,32]. To avoid the need for slow wavelength sweeping, the spectrally and spatially resolved interferometry (SSRI) technique was developed [33] through generating spectral interference patterns. However the broadband phase spectrum is measured in a single capture and a robust algorithm is needed to track the phase evolution through fringe patterns, to suppress artificial phase variations, as the conventional phase extraction algorithms are prone to phase noise [34–36]. To address these challenges, we developed an artificial intelligence (AI)-assisted algorithm to seamlessly extract quantitative phase spectrum and compute the phase variation value (PVV) for a diverse set of samples, enabling rapid, accurate, and stable broadband interferometry (see **Fig. 1b**). We demonstrate enhanced plasmonic metasurface phase sensing and hyperspectral single-cell dispersion imaging illustrating the capability and sensitivity of our proposed method over conventional spectroscopy (**Fig. 1c**). The presented experiments including mechanical perturbation, plasmonic metasurface sensing, and biological dispersion imaging, are not independent applications but validation cases demonstrating the method's robustness, relaxed polarization constraints, and broad applicability. Compared to existing common-path interferometry techniques, the proposed approach achieves an order-of-magnitude improvement in phase sensitivity while remaining versatile across diverse sample types.

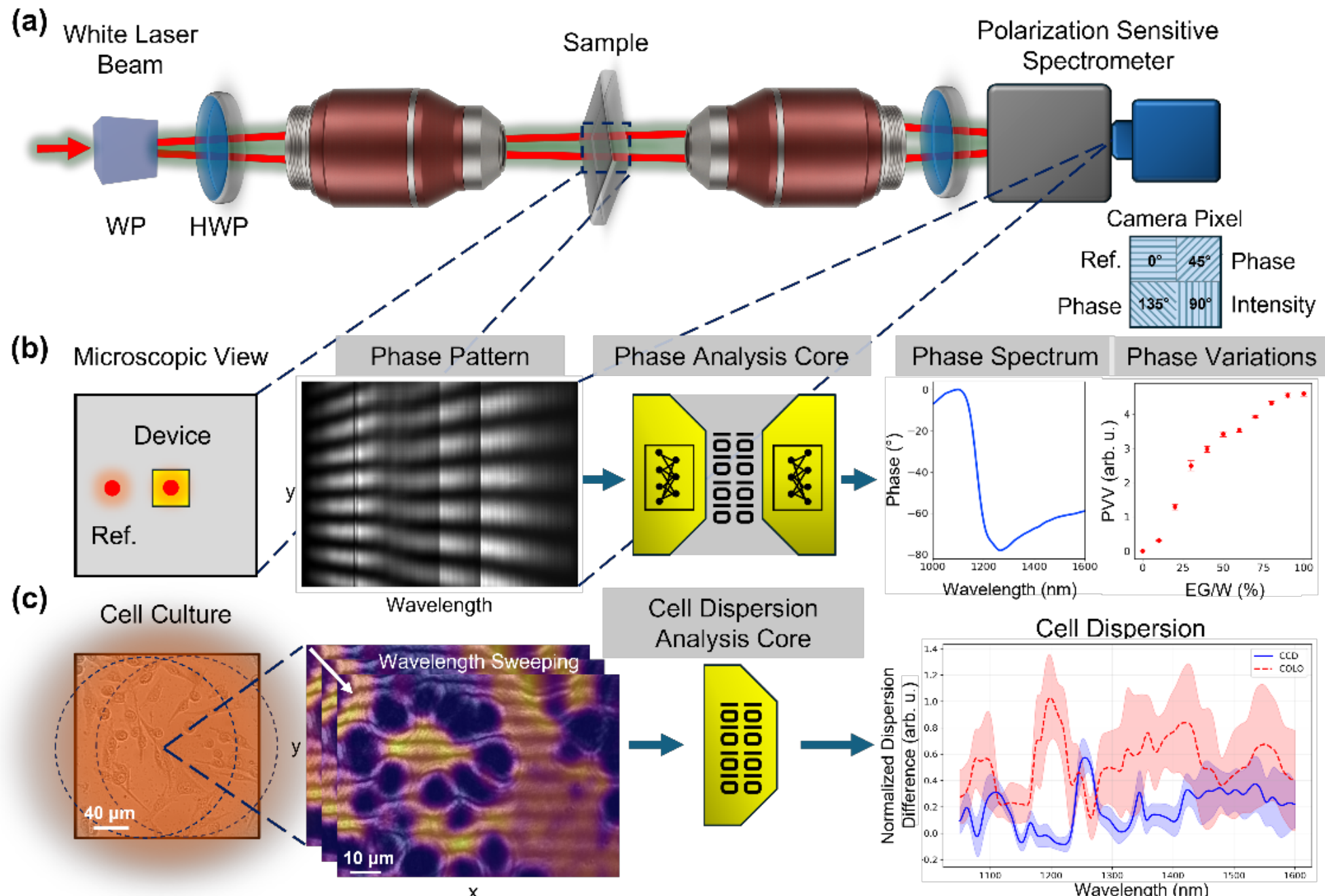

**Fig. 1.** GPCPI interferometry. (a) Simultaneous transmittance and phase spectra measurements via the reference and sensing beams formed by a Wollaston prism (WP). The amplitude and phase information can be extracted from different polarization angles, for instance the horizontal and vertical polarizations contain reference and sample intensities, while 45° and 135° angles form the interference pattern. The presented scheme increases the stability of the phase measurements, while relaxing polarization requirement of the sample, enabling a compact and robust broadband interferometer. Half-wave plates (HWPs) were used to control the polarization angle of the two beams. Details of the experimental setup are given in **Fig. S1-3**. (b) Phase extraction mode. Spectral interference pattern (phase pattern) formed by interfering reference and sensing beams is used to extract the quantitative phase spectrum of the sample. The extraction process is enhanced by using deep neural autoencoders for phase anomalies suppression. Furthermore, phase variation value (PVV) is estimated through post-processing the phase patterns using transfer learning-based trained ConvNeXt V2 model for AI-enhanced rapid phase sensing, enabling multiple applications such as biomarker detection. (c) Cell dispersion extraction mode. Illuminating a cell-cultured sample with reference and sensing beams forms a 2D interference pattern, containing the cell dispersion details. Analyzing spatial frequencies of the hyperspectral interference patterns reveals dispersion information of the cells, enabling label-free cell classification and disease diagnosis.

## 2. AI-enhanced optical phase sensing and measurements

### *2.1 GPCPI interferometry*

Conventional interferometry techniques (e.g., MI) require two separate optical paths, reference and sensing arms, to measure the quantitative phase of a sample, making them susceptible to phase noise induced by factors such as mechanical vibrations and environmental variations [31]. CPI addresses this issue by coalescing the two optical paths into one. However, it requires samples to have special polarization properties [21,22]. GPCPI has the low-noise and high stability of CPI method, while relaxing the polarization requirement. GPCPI divides a single broadband laser beam into two arms with orthogonal polarization using a Wollaston prism (WP). The two polarizations are then rotated to the desired polarization using a half-wave plate (HWP) (see **Fig.1a** and **Fig.S3**). The spectral interference pattern on the imaging spectrometer sensor (**Fig.1a** and **Fig.S3**) is formed by interfering the two orthogonal polarizations after passing through a polarizer at 45°. For simultaneous amplitude and phase measurements, a polarization sensitive camera, or double cameras with dedicated linear polarizers can be used (see, **Fig. 1a**). The almost common path of the two beams makes the GPCPI method significantly more stable than the MI method, while relaxing the polarization constraint of the CPI method. More details about optical setups of the above-mentioned interferometry techniques are provided in the Supplementary Materials.

### *2.2 Quantitative phase extraction*

The spectral interference pattern can be examined by Eq.1, considering small angle between the beams ($\theta \ll 1$):

$$I = I_1 + I_2 + 2\sqrt{I_1 I_2}\, cos(k\theta y + \Delta\varphi) \quad (1)$$

Where, $I_{1,2}$ are the beam 1 and 2 intensities, $k$ is the magnitude of the propagation vector, $y$ is the vertical component of the interference pattern (see **Fig.2e**), and $\Delta\varphi$ is the phase difference between the sensing and the reference arms (beams 1 and 2). A Fourier transform (FT) can be applied to Eq.1 to extract $\Delta\varphi$ from the phase pattern using experimental data and Eq.2:

$$\Delta\varphi = \tan^{-1}\left(\frac{Im\{FT(\omega_{+/-})\}}{Re\{FT(\omega_{+/-})\}}\right) \quad (2)$$

In equation 2, $\omega_{+/-}$ is the spatial frequency corresponding to the side peaks of the Fourier spectrum. The phase spectrum contains local (single wavelength) and global (multiple wavelengths range) discontinuities, that need to be processed to extract the actual phase spectrum of the sample (see **Fig.S4** and **S5**). More information is provided in the Supplementary Materials. To realize broadband phase measurements and address the limitations of the existing phase extraction algorithms, we introduced an AI-enhanced phase anomalies detection algorithm based on deep neural autoencoders (**Fig.2a-d**).

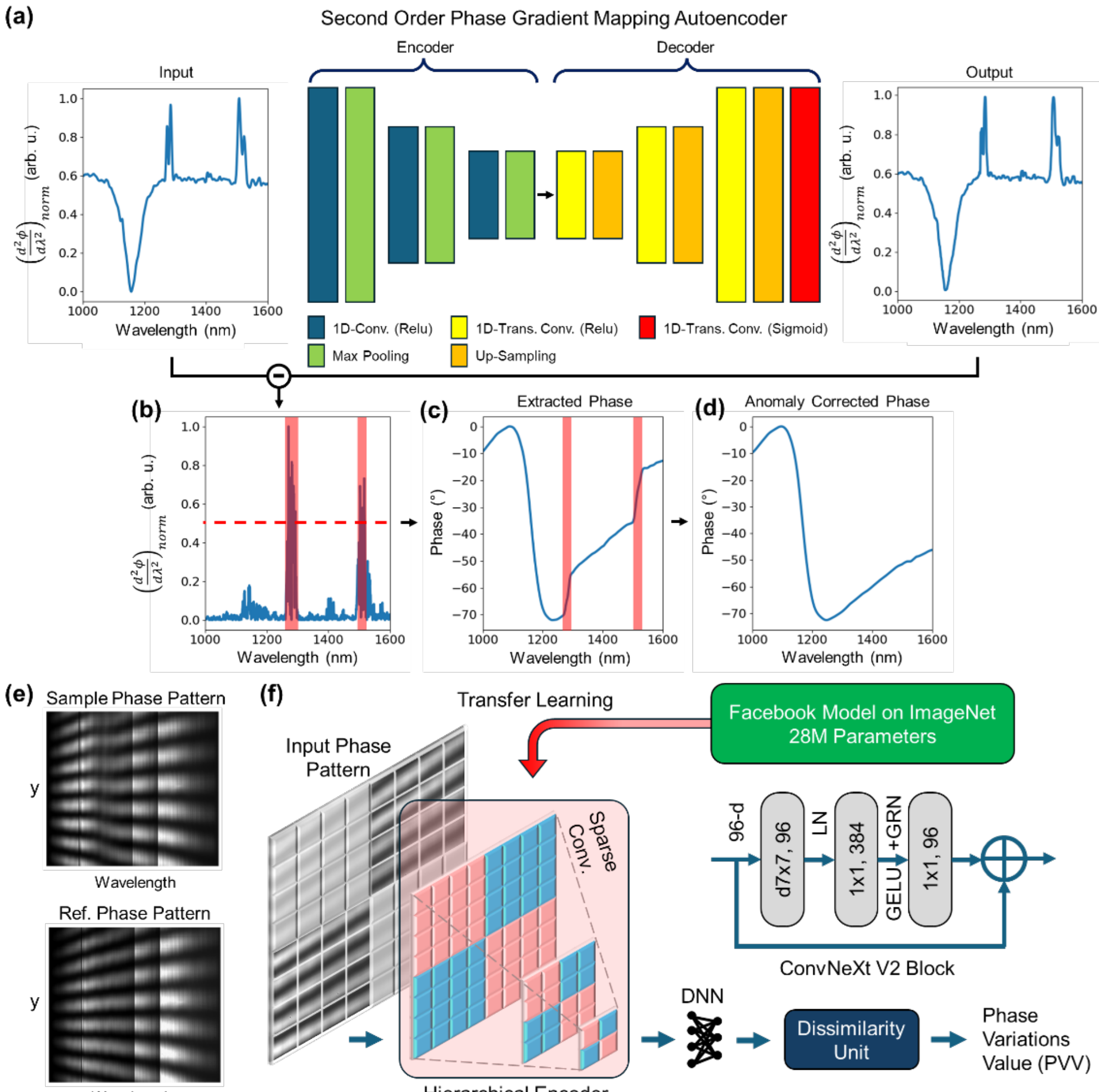

**Fig. 2.** AI-enhanced broadband phase sensing algorithms. (a) Deep neural autoencoder for phase anomalies detection using second order gradients. The network consists of series of convolutions to encode the input second order phase gradient spectrum into a compressed vector, then it regenerates the input by applying series of transversed convolutional operations. Note that the autoencoder weights were calculated through a training process with a dataset containing artificial anomalies inserted phase spectra. (b) Autoencoder phase spectrum anomalies detection. The autoencoder maps second order phase spectrum gradient to itself, resulting in anomalies peaking at phase residual spectrum (autoencoder error). (c) Detected phase residual peaks determine the location of phase anomalies in the calibrated phase spectrum. (d) Anomaly corrected phase forms by applying a global path-following technique, preserving global phase spectrum profile by correcting phase discontinuities at anomalies detected locations. The globally corrected spectrum exhibits smooth measured phase values for a plasmonic metasurface in the NIR region. (e) Sample and reference phase patterns. Note, the fringe lines in the sample phase pattern are shifted due to the sample induced phase change. (f) ConvNeXt V2 model for phase variations value (PVV) score estimation. Captured phase patterns are encoded into compact vectors by passing them through a hierarchical encoder and applying sparse convolutions to extract key features. The Facebook ConvNeXt V2 model was adopted for phase pattern dissimilarity analysis by fine-tuning the last layer of the deep neural network (DNN) using the transfer learning technique. Compared to basic CNNs, ConvNeXt V2 model offers the advantage of incorporating the global response normalization (GRN) mechanism inspired by vision transformers, enabling a global perspective of the input image, as illustrated in the ConvNeXt V2 block.

Our AI-assisted algorithm processes two phase patterns: the first, the reference phase pattern, is generated by passing two beams through the bare regions of the chip; the second, the sample phase pattern, is obtained by passing one beam through the bare region and the other through the sample (see **Fig. 2e**). Using the fast Fourier transform (FFT) technique and Eq. 2, these phase patterns are transformed into wrapped phase spectra, usually exhibiting several π discontinuities (see **Fig.S5**). To remove these artificial discontinuities, we employed a standard phase unwrapping technique that follows the phase spectrum path and vertically shifts the spectrum at discontinuities to align phase values before and after each discontinuity, a method known as path following [37]. Since this is applied locally (point-to-point), we refer to it as the local path-following algorithm. After unwrapping the sample and reference phase spectra (**Fig. S5**), the algorithm calibrates the sample phase by subtracting the reference phase, thereby removing any residual phase differences that exist even in the absence of the sample.

Despite the application of the local path-following technique, the calibrated phase spectrum may still contain anomalies due to the broadband nature of the measurements, which makes them sensitive to factors such as non-uniform spectral and spatial intensity distribution, intensity fluctuations, and dimmed imaging sensor pixels. To detect and correct these anomalies, we developed a neural network-based model by training a general autoencoder that converts the second order gradients of the smooth phase spectrum into their same copies using deep neural networks, comprising encoding and decoding units built with 1D convolutional neural networks (see **Fig. 2a**). The second order gradient helps to intensify the effect of discontinuities in the phase spectrum and its slope of variations. The convolutional operations compress the input spectrum into an encoded vector, which is then reconstructed by the decoding unit through a series of transverse 1D convolutional operations. By inputting into the autoencoder the measured phase spectrum second order gradient containing anomalies and calculating the binarized difference between the input and output, the locations of the anomalies are identified (shown as red, see **Fig. 2b**). More information about the anomaly detection algorithm is given in the Supplementary Materials. Additionally, we provided a manual option in the algorithm, allowing the user to select specific regions to be corrected based on prior knowledge, such as the positions of burned or dimmed pixels in the imaging sensor. Once all anomalies are detected, the algorithm aligns the phase values before and after each detected region and vertically shifts the rest of the spectrum (**Fig. 2c-d**). Since this path-following is performed over a region, we refer to it as global path-following.

### *2.3 Deep learning enhanced phase sensing*

Considering the phase sensitivity to nanometer (sub-wavelength) changes in the optical path length, phase sensing can detect minute changes such as refractive index variations in the sample. The spectral phase pattern carries rich information about both the phase and the relative intensity of the light passing through the sample. A deep learning model can extract encoded features from the captured interference patterns while excluding the noise contributing factors. We developed a neural network algorithm by training a noise-robust ConvNeXt V2 model [38], which enables hierarchical encoding of phase patterns and computation of the phase variation value (PVV) score (see **Fig. 2f**). PVV scores are calculated by comparing the encoded vectors of the perturbed and control (unperturbed) cases through dissimilarity computations using a cosine similarity algorithm. ConvNeXt V2 model represents an updated version of combined convolutional neural networks (CNNs) and residual networks (ResNet), incorporating the global response normalization (GRN) mechanism inspired by vision transformers. Since training the ConvNeXt V2 model requires a large dataset, we utilized transfer learning by adopting the Facebook ConvNeXt V2 model, which comprises 28 million parameters and is pre-trained on the ImageNet dataset. This feature provides a global perspective on the relationships between different parts of an input image, which is especially advantageous for phase pattern analysis, as it involves continuous phase information of the sample. Additional details about ConvNeXt V2 model and PVV score estimation are given in the Supplementary Materials.

## 3. Results and Discussion

### *3.1 Robust phase measurements*

Phase measurements are inherently sensitive to external perturbations. Therefore, perturbations such as mechanical vibrations generate phase noise in the conventional MI based techniques. To evaluate the phase stability of the proposed GPCPI method, a vertical shock was applied to the setup, and the phase pattern was recorded for 14 seconds for each of the interferometry techniques (see **Fig. 3**).

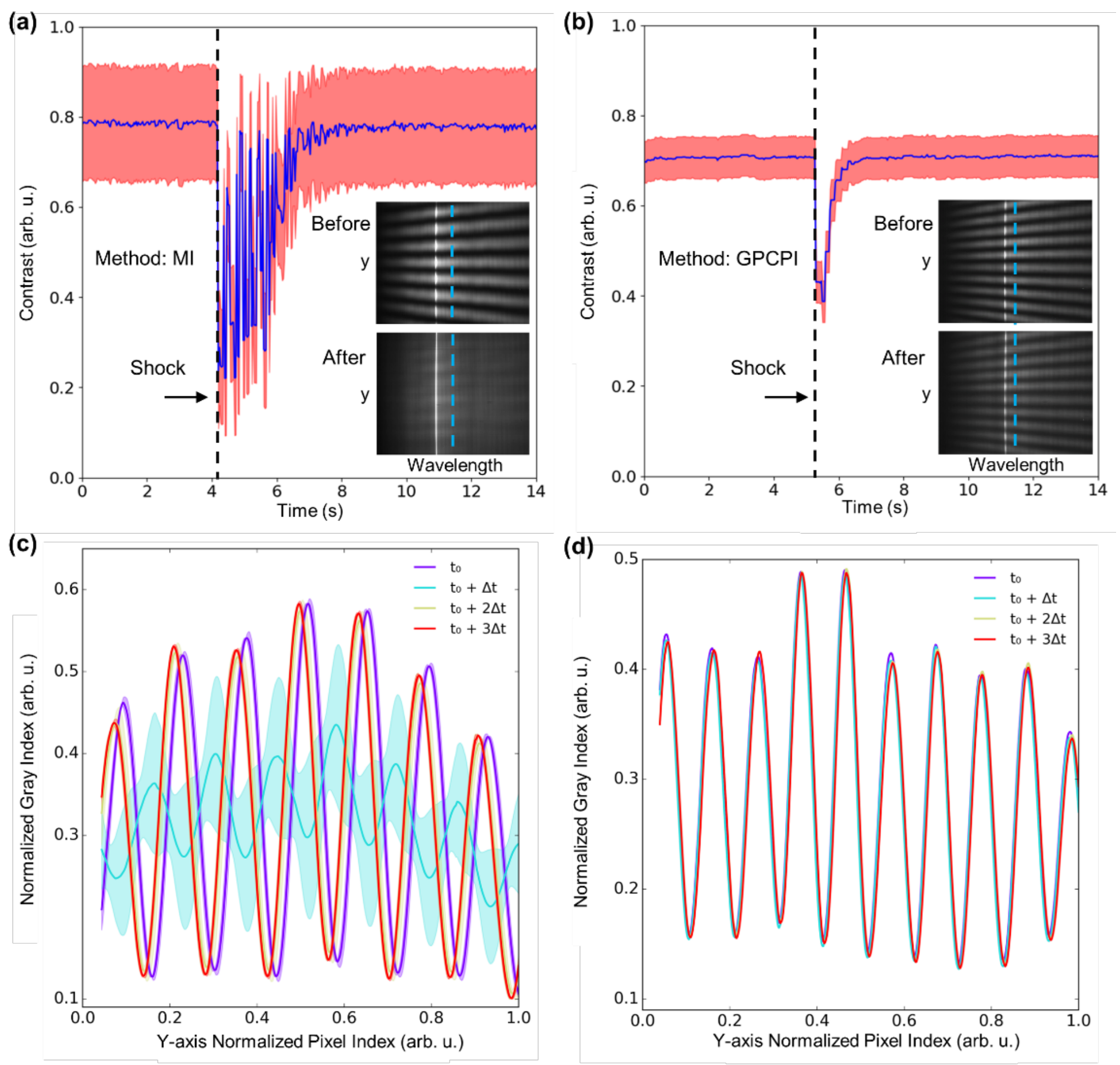

**Fig. 3.** Phase measurements stability analysis. (a-b) The normalized contrast of the phase pattern along the blue dashed line (see insets) over time for the MI and GPCPI methods, respectively. The blue solid line shows the mean fringe contrast, and the shaded red region represents the standard deviations. The results show large variations in the MI method, even before the shock was applied. In contrast, GPCPI exhibits smaller variations, leading to more stable phase extraction. The shock analysis reveals approximately 87% and 50% contrast reduction immediately after the shock and shock damping times of about 2.5s and 1.6s for the MI and GPCPI methods, respectively. The standard deviations at shock-free periods are approximately 31% and 14% for MI and GPCPI, respectively. Insets show projected interference patterns before and after a vertical shock was applied to the optical table. The interference pattern nearly vanished in the MI setup compared to the GPCPI pattern, where the pattern remained visible after the shock. (c-d) Vertical phase pattern variations along the blue dashed line at four consecutive time stamps ($t_0 = 0$ s, $\Delta t = 3.5$ s) for the MI and GPCPI methods, respectively. Solid lines show the mean fringe pattern, and the shaded area represents the standard deviations. The results indicate significant changes in the phase pattern in the MI method, both in magnitude and in the peak locations, whereas the GPCPI method maintains consistent pattern with minimal variations.

**Fig. 3a-b** presents the fringe contrast for GPCPI and MI interferometers after the application of a shock. A clear disturbance to the fringe contrasts is observed after the shock in the two interferometers. In Fig.3a-b, the blue plots represent the mean value of the fringe contrast along the blue dashed lines, and the red region shows the standard deviation of the contrast. The application of a shock significantly decreases the fringe contrast for the MI interferometer with 87% contrast drop, while the fringe contrast decreases by only 50% for the GPCPI interferometer. The magnitude of the standard deviation informs us about the passive variations in the interference pattern due to background vibrations. The larger standard deviations for the MI interferometer (~31%) compared to the GPCPI interferometer (~14%) is additional evidence of the improved stability of the later system. Additionally, the drifting of fringe patterns along the vertical blue dashed lines of **Fig. 3a-b** is presented in **Fig. 3c-d** for four consecutive times. The magnitude and peak locations of the cosine patterns of the interference fringes (**Fig. 3c-d**) evidences a significantly enhanced stability for the GPCPI technique (**Fig. 3d**) compared to the MI technique (**Fig. 3c**). The comparison with CPI method is presented in Supplementary Materials (see **Fig. S6**). The integration of the GPCPI improved stability and polarization flexibility with the AI-driven phase sensing enables a wide range of applications such as phase sensing and dispersion imaging that we will discuss in the next sections.

### *3.2 Application to metasurface-based refractive index sensing*

Measuring the phase spectrum of optical materials provides crucial information in evaluating the performance of photonics devices, validating electromagnetic computational models (e.g., local phase method [39]), and enhancing the sensitivity of bio-photonics sensors [20]. We first applied the GPCPI method to a plasmonic metasurface, where broadband measurements are important due to the lossy nature of the structures. The metasurface consists of an array of plasmonic nanorods designed to have a resonance around 1.2 μm. The metasurface is fabricated by electron-beam lithography (see **Fig. S7**) and encapsulated in a custom 3D-printed flow cell for liquid-based refractive index sensing (**Fig. 4a-b**). To perform phase sensing measurements, we flowed a mixture of water and ethylene glycol (EG) at different concentrations over the encapsulated metasurface. We varied the concentration from pure water to 100% EG in 10% steps, leading to bulk refractive index variations from 1.33 to 1.43 [40]. The flow cell simplifies liquid handling and enables the formation of a thin liquid film on top of the metasurface. The thickness of the liquid layer was kept below 50 μm to ensure low optical loss in the NIR spectrum due to water absorption.

We measured the transmittance of the sample under different concentrations of the ethylene glycol/water (EG/W) mixture to evaluate the refractive index sensing performance of the metasurface, comparing the results with simulated data (see **Fig. S8a**). The resonant frequency at each concentration was extracted using single-mode S-matrix fitting [41,42] to determine the resonance frequency shift in response to refractive index variations. Measurements show that the plasmonic sensor successfully detects a minimum concentration of 20% in our experiment, corresponding to a refractive index (RI) change of 0.02 (see **Fig. S8b**), relying on the transmittance measurements only. The designed plasmonic metasurface is a basic single-resonance sensor, with a sensitivity of ~ 1400 nm/RI. Note that the observed deviation from the simulated resonant frequency shift can be attributed to factors such as temperature variations, fabrication errors, misalignment, and non-normal illumination of the sample.

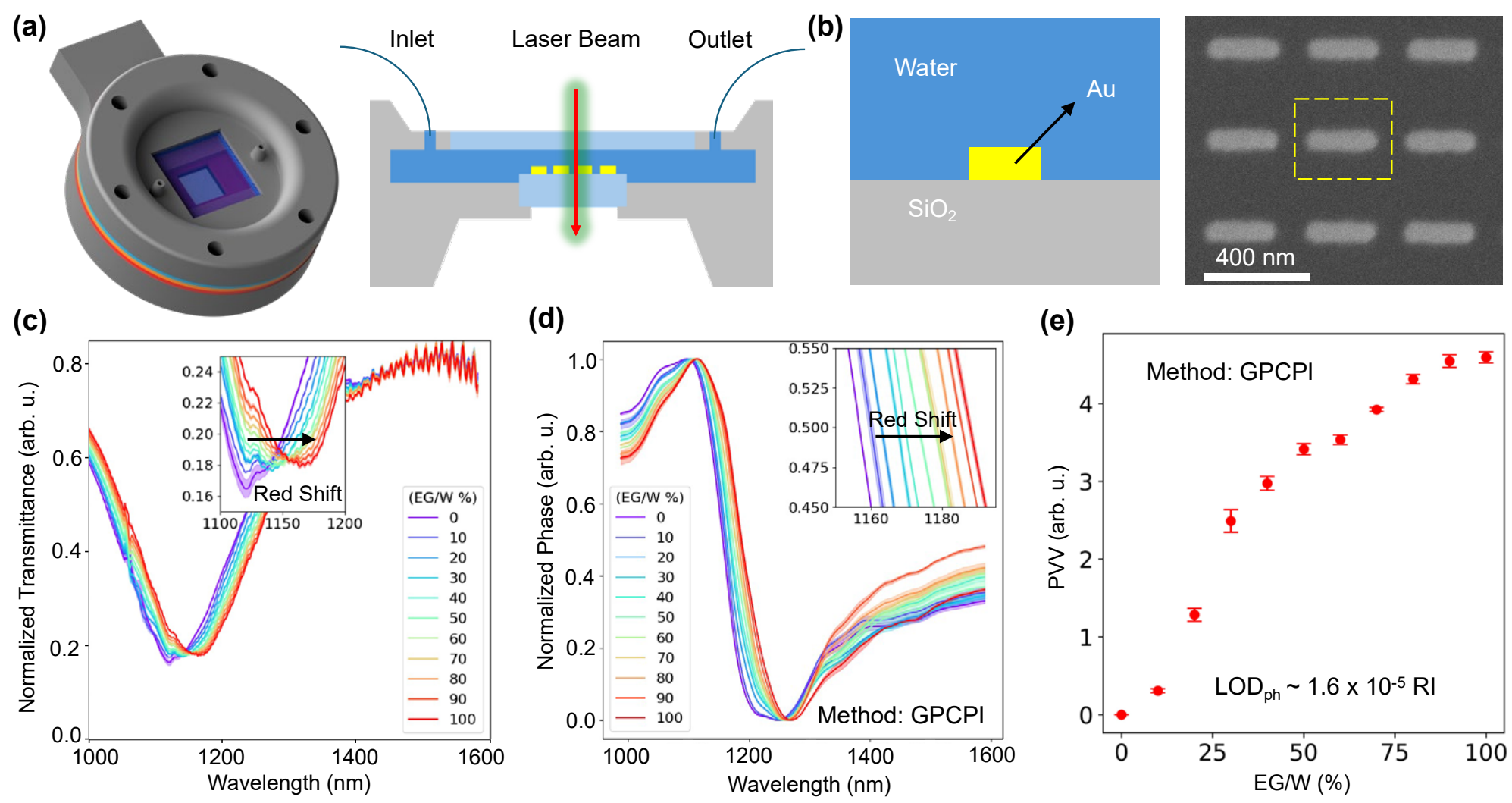

**Fig. 4.** Quantitative phase spectrum extraction and sensing mode. (a) Flow-cell integrated plasmonic metasurface chip for refractive index sensing. (b) Schematic of the unit-cell of a metasurface and the scanning electron microscope (SEM) image of a few unit-cells of a fabricate device. The metasurface consists of gold nano-bars with length, width, and thickness of 260 nm, 80 nm, and 40 nm, respectively, arranged in a 2D array with periodicities in x and y equal to 375 nm. (c) The experimental transmittance and (d) the broadband normalized phase spectra of the plasmonic metasurface at various concentration of ethylene glycol (EG) and water (W) mixture. Note, the experimental phase range is from 0° to -80°. GPCPI method was used to measure the metasurface phase spectrum. Refractive index variations from 0 to 0.1 were applied in the modeling to represent the EG/W mixture concentrations used in measurements. Note, each measurement was repeated three times to assess the stability of the phase sensing method. The minimum phase variation observed with the GPCPI method is approximately $1.75\times10^{-3}$°. Phase deviations between the measured and simulated spectrums caused mainly by the beam and device non-uniformities, temperature fluctuations, camera's quantum efficiency, and material properties deviations. (e) Refractive index sensing via PVV score monitoring of metasurface phase patterns under different EG/W mixture concentrations, showing enhanced sensitivity compared to the transmission-based sensing (**Fig. S8b**). Perturbed phase patterns were compared with the control (unperturbed) pattern by passing both images through the ConvNeXt V2 encoder and dissimilarity unit to estimate the PVV score (see **Fig. 2e-f**).

Apart from transmittance measurements, we captured the phase spectrum of the resonant metasurface at varying ethylene glycol/water (EG/W) concentrations (color-coded from blue to red, see **Fig. 4c-d**) to evaluate the phase variations under refractive index changes. Note that all interferometry techniques (MI, CPI, and GPCPI) along with our proposed algorithm were tested (**Fig. 4c-d** and **Fig. S9a-b**). **Fig. 4c-d** presents the measured transmittance and the phase spectra, evidencing a red shift as the concentration of the EG/W mixture increases. A good agreement between simulations and measurements is observed (see, **Fig. S8-9**). Note that spectral features of EG/W mixture do not affect the measured optical response of the metasurface as the measurements are referenced against device-free region of the chip. The stability of the phase measurement was assessed by analyzing its repeatability and a minimum phase variation of $\sigma_{ph} = 1.75\times10^{-3}$ ° for GPCPI method was observed, an order of magnitude improvement compared to the state-of-the art.

To facilitate the phase variation tracking, we applied the ConvNeXt V2 model to phase patterns obtained from various interferometry techniques, demonstrating robust refractive index sensing (see **Fig. 4e** and **Fig. S11**). The resulting PVV values were monitored to detect the levels of applied perturbations without directly extracting individual phase spectrum, effectively minimizing noise-contributing factors and the readout time, thus enabling real-time perturbations monitoring. PVV scores show enhanced sensing compared to the transmission-based method, capturing variations as small as 0.01 RI in the surrounding medium (minimum prepared EG/W mixture). Considering the steepest phase variations part of the spectrum with applied refractive index change, our system shows phase change per refractive index, $S_{ph} \sim 330$ °/RI (a basic plasmonic resonator). The phase-based calculated limit of detection ($LOD_{ph}$) of the current systems is $LOD_{ph} = 3\sigma_{ph}/S_{ph} \sim 1.6 \times 10^{-5}$ RI. Note, the phase variations with refractive index can be quantified by calculating the accumulated phase difference across the measured spectra (see **Fig. S10**), however it suffers from phase noise and depends on the spectral integration range, highlighting the improvement brought by PVV monitoring. Applications of the proposed interferometry method are not limited to phase sensing alone and we present dispersion imaging of biological samples, providing rich information for cell classification and disease diagnosis.

### *3.3 Application to cell dispersion imaging and classification*

Label-free biological sensing enables rapid and non-invasive molecular tracking within biological cells [43]. Recently, label-free quantitative phase imaging was used to track protein aggregation at the molecular-level using the phase information of incident light [44]. Although the phase imaging showed protein aggregation detection, it works mostly based on the geometrical clues of the captured images without recording the spectral information. Our interferometry technique enables molecular fingerprinting through broadband dispersion imaging as shown in **Fig. 5**. To capture the dispersion profile over the wide spectral band, we used acousto-optic modulator, sweeping the beam's central wavelength from 1050 nm to 1600 nm in NIR range, using hyperspectral imaging technique. Compared to the phase sensing mode, in the dispersion imaging, both the reference and sensing beams illuminate the sample (e.g., cell-cultured) where the beams wave-vectors get perturbed by the dispersion and the topology of cells (see **Fig. 5a**). To illuminate the sample uniformly, both beams were focused at the back focal plane of the condenser lens. As we capture the interference pattern over the imaging sensor, the relative angle between the rays is the only parameter that contains the information of the sample (see Eq.1). This angle increases particularly at the boundaries of the sections with different refractive indices, resulting in the deformation of the fringe patterns (**Fig. 5a**). These deformations translate to the local variations in the fringes pattern frequencies, enabling cells optical dispersion imaging and classification.

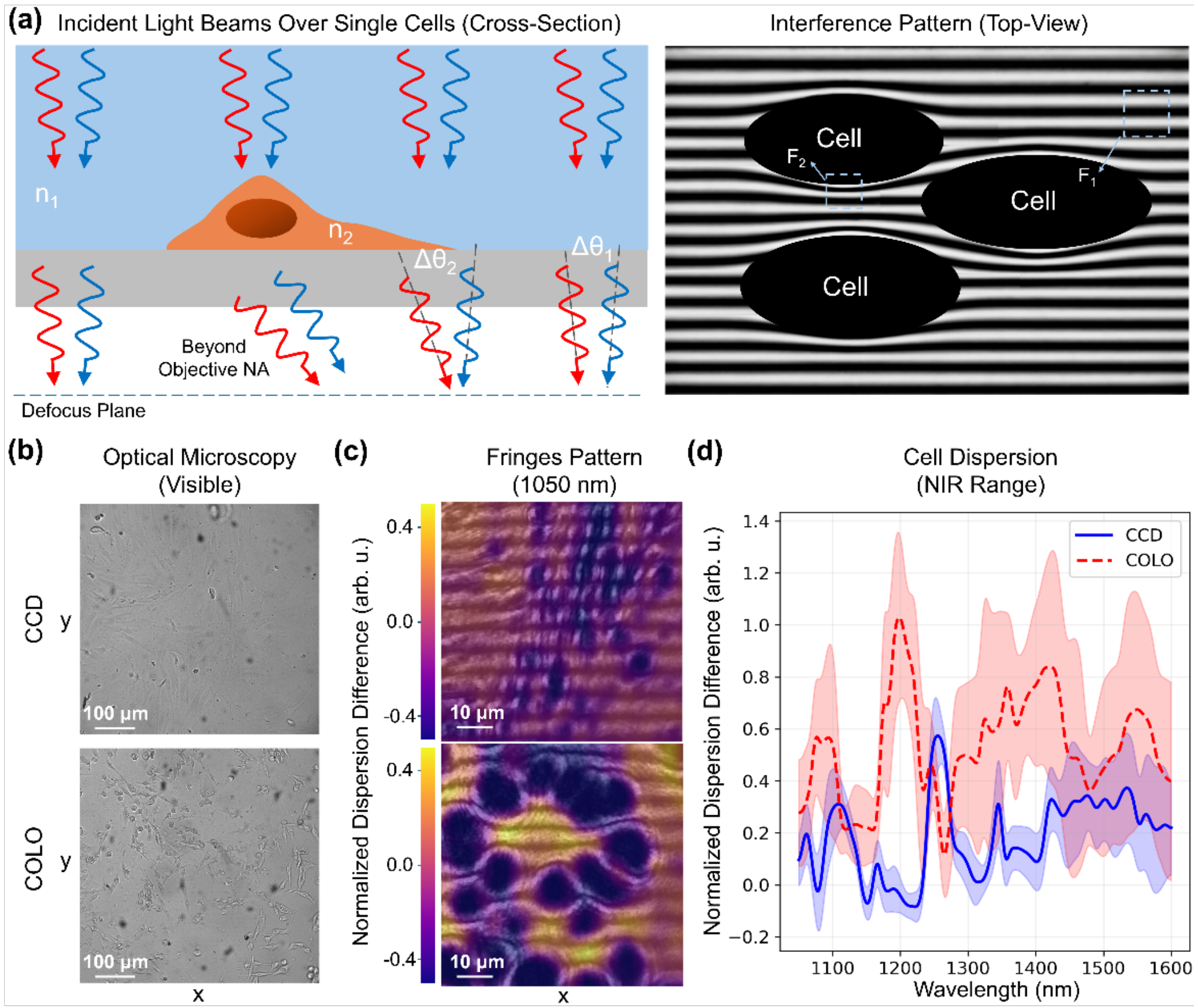

**Fig. 5.** Cell dispersion imaging mode. (a) The reference and sensing beams passing through the cell-cultured sample, generating a spatial frequency distribution across the 2D plane after interfering over the imaging sensor. The combined effect of the cell topology and its refractive index perturb the incident beam wave-vector, causing fringes pattern deformation. The interference pattern captured by placing the sample at the defocus plane to increase the fringes deformation while capturing the individual cells. Analyzing the fringes pattern spatial frequencies reveals the optical response of cells, containing cells specific information. A hyperspectral technique was used to collect interference patterns at multiple wavelengths. (b-c) Optical microscopic image and the fringes pattern (at 1050 nm) of CCD and COLO skin cells, respectively. The interference pattern was collected over a broadband range (1050–1600 nm) in the NIR spectrum using a tunable acousto-optic filter. The normalized dispersion difference map is overlaid on top of the interference fringes. In the NIR range, CCD cells have a negligible effect on the fringes pattern spatial frequency whereas COLO cells significantly deform the interference fringes. (d) Estimated cell dispersions of CCD and COLO skin cells through calculating the normalized dispersion difference. Captured spatial frequencies solely affected by the cells through perturbing the relative angle between rays (see Eq. 1). Cell dispersion profiles reveal unique fingerprints that enable classification of different cell types and support disease diagnosis. Note that the dispersion profiles averaged over 15 areas on the cell-cultured sample to minimize the cell-to-cell deviations.

To show the capability of our method, we tested normal versus cancerous skin cells, called CCD (CCD-32Sk cell line) and COLO (COLO-829 cell line), respectively. Conventional optical microscope images of the cells mostly evidence the geometrical features, while the cell dispersion information is missing (**Fig. 5b**). However, interferometric imaging reveals the dispersive response of cells, providing the cell-specific fingerprints in the NIR. Comparing the normal versus cancerous cells, the captured interference fringes in the CCD case remain approximately untouched while cancerous cells causing significant deformation due to the stronger combined effect of COLO cells topology and dispersion (see **Fig. 5c**). To amplify fringes deformations, we placed the sample slightly off the focus plane, within a range that cells are still visible, so the relative angles between the rays increases at the interference plane. Note that cells imaged at the defocused plane are seen as dark objects since the refracted beams at cells locations are outside of the objective lens acceptable numerical aperture (NA).

To process the hyperspectral interferometric images, we applied FFT locally to convert the fringes pattern into the frequency map of the sample where the optical phase variations of the incident beams can be retrieved. The conventional quantitative phase imaging technique outputs the overall phase variations by the sample [10], however, the dispersive response can be retrieved by multi-wavelength scanning of the sample and applying adequate phase normalization [45]. Hence, to decouple the dispersive response from the geometrically affected part, we calculated the normalized dispersion difference of the sample versus the background medium. Considering the small angle between incident beams (~ 1°) and cells as pseudo plano-convex micro lenses, the dispersion difference between the cell and the background medium is proportional to the spatial frequency differences between these two areas. Details of the proposed method are explained in the supplementary materials. Note, the geometrical effect of the cell was approximated by the lens maker formula, where the refractive power of the lens depends on the ratio of the refractive index contrast over the radius of the curvature. The calculated spatial frequency difference was normalized against the computed frequency at the unperturbed area to cancel out the effects of the defocus variations among different samples and the measurements wavelength, resulting in the normalized dispersion difference factor (see **Fig. 5c**). It should be noted that mapping the normalized dispersion difference on to the captured interferometric patterns enables single-cell dispersive imaging of the sample (**Fig. 5c**), showing the individual cell refractive response. Furthermore, we tracked the normalized dispersion difference over broadband wavelength range (1050 nm to 1600 nm), through application of the proposed method at each wavelength (see **Fig. 5d**). Comparing the CCD versus COLO cells (blue solid and red dashed lines, respectively, **Fig. 5d**), COLO's normalized frequency behaves differently, providing information for cancerous versus normal cell classification. The statistics of the estimated dispersions were obtained by averaging over 15 perturbed fringes areas (shown as shadow region in **Fig. 5d**) of a sample having over 10,000 cells. It should be noted that a medical grade analysis requires averaging over diverse set of cells from wide range of patients, however, the observed difference between normal and cancerous cells dispersions in our study shows the capability of the proposed method in cell classification and disease diagnosis.

## 4. Conclusion

We introduced general polarization common-path interferometry (GPCPI), a compact, robust, and broadband simultaneous amplitude and phase spectra measurement technique by relaxing the polarization constraint of the conventional common-path interferometer. The introduced technique is enhanced by an AI-assisted algorithm for accurate and real time polarization decoupled amplitude and phase measurements. GPCPI captures the phase spectrum of arbitrarily polarized samples while maintaining identical environmental conditions for both sensing and reference beams, resulting in over an order of magnitude enhancement in phase measurement stability. We applied GPCPI to characterize the phase response of a plasmonic metasurface under refractive index variations, demonstrating improved sensing performance. Real time noise suppressed phase sensing was shown by training ConvNeXt V2 deep neural model on captured interference patterns. Beyond phase sensing, we presented hyperspectral cell dispersion imaging by spatial frequency analysis of interference fringes, enabling cell classification and disease diagnosis. In summary, the GPCPI method is versatile and broadly applicable to metrology, optical material assessment, molecular diagnostics, drug discovery, and quantum sensing.

**Funding.** NIH S10OD034382 and ARO W911NF2310027.

**Disclosure.** The authors declare no conflicts of interest.

**Data Availability.** Data supporting the findings of this study are available within the paper and from the corresponding author upon reasonable request.

**Supplemental Document.** See Supplementary Materials for supporting content.